
\magnification=1200
{\nopagenumbers
\baselineskip=20pt
\hsize=5.5in
\hskip10cm UM--P-94/50\hfil\par
\hskip10cm RCHEP--94/13\hfil
\vskip2cm
\hskip2cm{\bf
  CP Violating Effects in $e^+e^- \rightarrow ZH$ and Their
 Optimization} \par
\vskip2cm
\hskip4cm J. P. Ma and B. H. J. McKellar\par
\vskip0.5cm
\hskip4cm Recearch Center for High Energy Physics \par
\hskip4cm School of Physics\par
\hskip4cm University of Melbourne \par
\hskip4cm Parkville, Victoria 3052\par
\hskip4cm Australia \par
\vskip2cm
{\bf\underbar{Abstract}}:
We consider the possibilities of testing CP conservation
in the process $e^+e^- \rightarrow ZH$.
The easiest is to measure the forward-backward asymmetry of the
$Z$ boson, a nonzero value of which signals CP violation. The effects of CP
violation are predicted in the two Higgs doublet model.
We show here that a choise of optimal observables can improve the possibility
of observing CP violation.
Although the number of  events is
not large, useful information about CP violating
parameters can be extracted.
\par\vfil
\eject}
\pageno=1
\baselineskip=20pt
{\bf 1. Introduction} \par
After its discovery in the Kaon system 30 years ago[1], CP violation is
still a puzzle. The standard
model can  accomodate CP violation in the Kaon system, where CP violation
is introduced through  the imaginary part of the CKM matrix elements. However,
because of the special structure of the CP violating interaction[2]
in the
standard model,
predicted CP violating effects in high energy processes, such as
high energy scattering, are extremely small. An explicit example is
given in [3]. Hence any CP violation found in high energy processes
will indicate the new physics beyond the standard model.
In any case, the standard model is generally regarded as a low energy effective
theory which will be modified or extended at high energies.
One of the possible extensions is
the two Higgs doublet model with two Higgs doublets instead of one.
In this model CP violation can
arise not only from the complex CKM matrix elements, but also from the
Higgs sector[4], where, for example, the neutral Higgs bosons with different
CP eigenvalues can mix. CP violation from this mixing may have
sizeable effects in high energy scattering processes, which are
studied in [5,6,7,8]. This model also allows a
significant electric dipole
moment for elementary particles[9]. \par
In this work we will study possible CP violation from the two Higgs doublet
model in the process $e^+ e^- \rightarrow Z H$, where the initial beams
are unpolarized. In this process it turns out that a CP test is possible
without the knowledge about the polarization of the
final state. A nonzero
forward-backward asymmetry of the $Z$ or $H$ boson indicates CP violation.
This is in contrast to the case of  $e^+e^- \rightarrow {\rm
 particle + antiparticle}$, where a CP test is only possible when one
measures the polarizattion of the final state. However, we will also
consider the possible information about CP violation revealed by
 the polarization
of the $Z$ boson, as the leptonic decay of $Z$ can measure its
polarization.
\par
In this work we also consider the optimization of CP violating effects. In
experiment, if the number of events avaliable for CP test is not
large, the statistical error can be
the main source of
error in the observation.  It is possible to reduce
the statistical error by a suitable
optimization procedure.
A general method for optimization was proposed recently[10,11].
We apply the procedure to the present process and compare
the effects  with and without optimization.
\par
We begin in Sect. 2 with our general formalism and introduce
some simple CP odd observables. In Sect. 3 we estimate the effect
of CP violation from the two Higgs doublet model and the  sensitivity of
our simple CP odd observables to CP violating parameters in the model.
In Sect. 4 we apply
 the optimisation
procedure mentioned above. Sect. 5 includes a discussion and
our conclusion .
\par
CP violation in $e^+e^- \rightarrow Z H$ is also discussed in [12]
in relation to CP violation in the Higgs decay, where only the real part
 of possible CP violating form factors is considered. In this case
a nonzero forward-backward asymmetry can not be produced.
\par\vskip 20pt
{\bf 2. CP odd observables}\par
We consider the following process with the unpolarized initial state:
 $$ e^+ (p_+) +e^-(p_-) \rightarrow Z(k) +H(k_H) \rightarrow
 \ell^+ (k_+)+ \ell^- (k_-) + H(k_H)  \eqno (2.1) $$
The momenta of the particles in the C.M.S. of the initial state
are indicated in the brakets. The lepton pair comes form the $Z$
boson decay. At the tree level, the Higgs boson in the process (2.1)
is produced through Bremsstrahlung off the virtuel
$Z$ boson produced by $e^+e^-$ annilation in the standard model.
\par
Instead of the momentum $k_+$ in Eq. (2.1), we will use
the momentum $q_+$ of $\ell^+$ in the $Z$
rest frame, which is of course related to ${\bf k_+}$
by a Lorentz boost.
We use
the notation ${\bf\hat k}$ to indicate a unit vector in the direction of
${\bf k}$.
We can introduce a density matrix
$R({\bf p_+}, {\bf k}, {\bf q_+})$ for the process in Eq. (2.1),
which we normalize so that
 $$ {\sigma(e^+e^-\rightarrow ZH \rightarrow \ell^+ \ell^- H)
   \over {\rm Br}( Z\rightarrow \ell^+ \ell^-) }
 = {1\over 2s}\cdot {\vert {\bf k}\vert \over 4\pi \sqrt{s}}
 \cdot {1\over 4\pi} \int d\Omega \cdot {1\over 4\pi}\int d\Omega_+
   R({\bf p_+}, {\bf k}, {\bf q_+}) \eqno(2.2) $$
Here $s=(p_++p_-)^2$ and $d\Omega (d\Omega_+)$ is the solid angle of
the vector ${\bf k}({\bf q_+})$. The quantity $R$ can always be written
in the form:
 $$ R({\bf p_+}, {\bf k}, {\bf q_+}) =A({\bf p_+}, {\bf k})
   + {\bf\hat q_+}\cdot {\bf B}({\bf p_+}, {\bf k})
   +(\hat q_{+i}\hat q_{+j}- {1\over 3} \delta_{ij})
     C_{ij}({\bf p_+}, {\bf k}) \eqno (2.3) $$
If CP invariance holds, we have:
 $$ A({\bf p_+}, {\bf k})=A({\bf p_+}, -{\bf k}), \ \
    {\bf B}({\bf p_+}, {\bf k})={\bf B}({\bf p_+}, -{\bf k}), \ \
  C_{ij}({\bf p_+}, {\bf k})=C_{ij}({\bf p_+}, -{\bf k})
  \eqno(2.4) $$
To test CP conservation one therefore has to verify
the equations in Eq. (2.4).
In Eq. (2.3) the quantities
${\bf B}$ and $C_{ij}$ contain the information about the polarization
of the $Z$ boson.  Without measuring the $Z$ polarization one
can still test CP
invariance by checking the CP constaint for $A$ in Eq. (2.4).
We propose the following CP odd observables for a CP test:
 $$ O_1=x={\bf\hat p_+}\cdot {\bf\hat k}, \ \ O_2=( {\bf\hat q_+}\cdot
   {\bf\hat p_+})({\bf\hat q_+}\cdot{\bf\hat k}), \ \
   O_3=({\bf\hat q_+}\cdot {\bf\hat p_+}) {\bf\hat q_+} \cdot
   ({\bf\hat p_+}\times {\bf\hat k}) \eqno(2.5)$$
A nonzero expectation value of any observable $O_i$ in Eq.(2.5) signals
CP
violation. We will call these observables simple CP odd observables.
Note that the $< O_1 >$ can only be nonzero if the quantity $A$ does
not satisfy its CP constraint. The forward-backward asymmetry of the
$Z$ boson can be
defined through  $O_1$:
   $$ a_{CP} = < \theta (O_1)- \theta (-O_1) > =
    { N(O_1 > 0)-N(O_1 < 0) \over N(O_1 > 0)+ N(O_1 <0) }
  \eqno(2.6)$$
where $N(O_1 > 0)$ and $N(O_1 < 0)$ is the number of events number
with $O_1 >0$ and $O_1 < 0$ respectively.
\par
\vskip 20pt
{\bf 3. CP violation from the two Higgs doublet model}\par
We consider here a two Higgs doublet model as a possible extension
of the standard model. In this model CP can be violated
 due to the complex
expectation values of the Higgs doublets. However, in general CP
violation occures together with the presence of flavour changing neutral
currents at the tree level. The flavour changing neutral currents can
be eliminated by imposing some discrete symmetry on the model. This
dicrete symmetry is softly broken in the Higgs potential[4]. In this
way it is possible to have CP violation in the model with the absence
of the flavour changing neutral currents at the tree level.
\par
In this model, there are three neutral Higgs bosons, two of them have
the CP eigenvalue $+1$, and the other one has $-1$ as its CP eigenvalue,
if CP invariance holds. In this case,  the neutral Higgs boson with
the CP eigenvalue $ -1$ can not be produced in the process (2.1) at the tree
level, because it decouples from two $Z$ bosons.
If CP is violated, the neutral Higgs bosons do not have any definite
CP eigenvalue, but are mixtures of the CP eigenstates.
In the process we consider
the effect of CP violation can occur at one loop level only
through the type of the diagrams given in Fig.1, where the loop is a
fermion loop. If the mass of the light flavours is neglected
compared to $m_t$, only
the top quark contributes.
In the two Higgs doublet model we write the relevant interaction
between the neutral Higgs boson $H$ and the top
quark and the interaction between two $Z$ boson and the Higgs boson $H$ as:
   $$ L_{Ht\bar t}= -iB_t \bar t \gamma_5 tH,
    \ \   L_{ZZH}={eM_Z\over 2\sin \theta_W \cos \theta_w }d\cdot
       Z^\mu Z_{\mu} H \eqno(3.1)$$
The parameters $B_t$ and $d$ are real. Furthermore $\vert d \vert <1$.
If $B_t$ and $d$ are simultaneously nonzero, CP is violated.
\par
We parameterized
the CP violating part of the $VHZ$ vertex with a form factor $F_V$ as
  $$ i\Gamma_V^{\mu\nu}(k,q)= {ie \over 2 \sin\theta_W \cos\theta_W}
       \epsilon^{\mu\nu\sigma\rho}k_\sigma q_\rho F_V(q^2)
      \eqno(3.2) $$
where the indices  $\mu$ and $\nu$ are
the Lorentz indices for the polarization
of the virtual vector boson $V$ and the real $Z$ boson respectively. $V$
stands for a virtual photon or a virtual $Z$ boson and has the momentum
$q$ with $q^2=s$, where  $s > (M_H+M_Z)^2$.
The on-shell $Z$ boson carries the momentum $k$. Calculating the relvant
diagrams, one finds that the real part of
the form factor $F_V$ has a complicated form, expressed
with Spence functions. Since we will take $\sqrt{s}=\sqrt{q^2}=500$GeV,
and the mass of the top quark will be taken as $175$GeV, as suggested by
recently in experiment[13], the absorbtive part of the vertex, i.e.
the imaginary part of the form factor, is always nonzero in the $s$ channel.
We calculate the imaginary part first, the real part can then be obtained
from the dispersion relation. For $V=Z$ the imaginary part is:
   $$ \eqalign {&{\rm Im} F_Z(s) ={6 e m_t B_t \over
      \sin\theta_W \cos\theta_W }\{ (g_V^t)^2 I_V(s) +
        (g_A^t)^2 I_A(s) \} \cr
      & I_V(s) ={D_L \over 16\pi\beta_Z E_Z\sqrt{s}}, \ \
      I_A(s)= {E_Z-\sqrt{s} \over 16\pi\beta_Z^3  E_Z^2 s}
      \{ (\sqrt{s}-E_Z+\beta_Z^2 E_Z) D_L
        +2\beta_t\sqrt{s} \} \cr
   & D_L =\ln { (s-M_H^2-M_Z^2) -(s-M_H^2+M_Z^2)\beta_t\beta_Z
          \over (s-M_H^2-M_Z^2) +(s-M_H^2+M_Z^2)\beta_t\beta_Z }
     \cr }\eqno(3.3) $$
Here $\beta_t$ and $\beta_Z$ are the velocity of the top quark and
the $Z$ boson in the rest frame of the virtual $Z$, and $E_Z$
is the energy of the real $Z$ boson.
${\rm Im} F_\gamma (s)$ can be obtained from Eq. (3.3) by the substitution:
    $ (g_V^t)^2 \rightarrow {4\over 3}\sin\theta_W\cos\theta_W g_V^t$ and
    $(g_A^t)^2\rightarrow 0$. From the imaginary part we can obtain the
real part:
   $$ {\rm Re}F_V(s) ={1\over \pi} P\int_{s_0}
      ^\infty d s' {{\rm Im} F_V(s') \over s'-s}, \ \
     s_0={\rm max}\{4m_t^2,\ (M_H+M_Z)^2\} \eqno(3.4) $$
Note that there is also another contribution to the imaginary part
of the form factor, if the mass of the Higgs boson is larger than $2m_t$.
We will not consider this case.
The CP violating part of the density matrix $R$ is generated by
the interference between the amplitude at tree level and the amplitude
including the vertex in Eq.(3.2). The result for $R$ is:
  $$ \eqalign { R(&{\bf p_+},{\bf k}, {\bf q_+}) =
   \left ( { e d \over \sin^2\theta_W\cos^2\theta_W}
           \right )^2 {3 s\over 64(s-M_Z^2)^2}
    \bigg\{ \lbrack E_Z^2 (1-\beta_Z^2x^2)
      + M_Z^2 ({\bf\hat p_+}\cdot {\bf\hat q_+ })^2\cr
     &    +2M_Z(E_Z-M_Z) x({\bf\hat p_+}\cdot {\bf\hat q_+ })
       ({\bf\hat k}\cdot{\bf\hat q_+ })
      + ({\bf\hat k}\cdot {\bf\hat q_+ })^2(-\beta_Z^2E_Z^2
     +x^2(E_Z-M_Z)^2 \rbrack\cr
     &+{4E_Z\sqrt{s} \over M_Z}\beta_Z\sin\theta_W\cos\theta_W
       (1-{ M_Z^2 \over s} ) {1\over d} {\rm Im}F_\gamma (s)
     \lbrack ({\bf\hat p_+}\cdot {\bf\hat q_+ })({\bf\hat k}\cdot{\bf\hat q_+
})
      M_ZE_Z \cr
      & +({\bf\hat k}\cdot {\bf\hat q_+ })^2 xM_Z(M_Z-E_Z)
      +xM_Z^2 \rbrack \cr
     &  -{ \sqrt{s} \over  d} {\rm Re} F_Z(s)
       \beta_Z E_Z {\bf\hat q_+}\cdot (
        {\bf\hat p_+} \times {\bf\hat k}) \lbrack M_Z
         ({\bf\hat p_+}\cdot {\bf\hat q_+ })
      +x(E_Z-M_Z) ({\bf\hat k}\cdot {\bf\hat q_+ })\rbrack \bigg\}
    \cr } \eqno(3.5) $$
Here we used the fact that the coupling constant $g_V^\ell (=-{1\over 2}+
  2\sin^2\theta_W)$ of the
$Z$ boson to the  neutral vector current is much smaller than
the coupling constant $g_A^\ell$ to the neutral axial current and neglected
$g_V^\ell$. The CP conserving part of $R$ at the tree level is also
given in Eq. (3.5). With the density matrix we can now calculate the
expectation value of the simple CP odd observables and the CP asymmetry in
Eq. (2.6). These quantities are proportional to $ {B_t \over d}$:
    $$ <O_i>=w_i {B_t\over d}, \ \ \, a_{CP}=y{B_t \over d} \eqno(3.6) $$
Further, $<O_{1,2}>$ receive contributions only from the imaginary part
of the form factors, while the real part contributes to $<O_3>$.
We take $\sqrt{s}=500$GeV and $m_t=175$GeV and obtain the numerical
results for the coefficients in Eq. (3.6):
  $$ \eqalign { w_1&=5.6\times 10^{-3},
       \ \ \ w_2=3.8\times 10^{-3},\ \ \
        w_3=-4.4\times 10^{-4},\ {\rm for }\ M_H=100{\rm GeV}\cr
        w_1&=6.7\times 10^{-3}, \ \ \  w_2=4.3\times 10^{-3}, \ \ \
        w_3=-4.6\times 10^{-4},\ {\rm for}\ M_H=200{\rm GeV} \cr
       } \eqno(3.7)$$
For the density matrix $R$ given in Eq. (3.5), $y$ has a simple
relation to $w_1$: $y={3\over 2 }w_1$.
The result for $y$ as a function of the Higgs mass $M_H$ is plotted
in Fig.2.
{}From Fig.2 note that $y$ increases as $M_H$ increases, mainly due to
the decreasing cross section with the increasing Higgs mass.
\par
To study how sensitive our observables are
to the CP odd parameter $ {B_t \over d}$,
we introduce the sensitivity $\rho_i$
of a observable $O_i$:
   $$ \rho_i={\sqrt{<O_i^2>}\over \vert w_i\vert} \eqno(3.8) $$
Remembering that the statistical error of a observable $O_i$ is given by:
   $$ \delta <O_i> = \sqrt{ {<O_i^2> -<O_i>^2 \over N}}
   \approx \sqrt{ {<O_i^2> \over N}} , \eqno(3.9) $$
where $N$ is the number of the avaliable events for measuring the
observable $O_i$, the sensitivity $\rho_i$ has the meaning that
the parameter ${B_t\over d}$ must be at least larger than
 ${\rho_i \over\sqrt{ N}}$ to have a measurable effect on $O_i$.
The results for $\rho_i$ are:
  $$ \eqalign{
       \rho_1&=88, \ \ \ \rho_2=80,\ \ \ \rho_3=558 \ \ \
    {\rm for }\ M_H=100{\rm GeV}\cr
      \rho_1&=75, \ \ \ \rho_2=72, \ \ \ \rho_3=532 \ \ \
    {\rm for}\ M_H=200{\rm GeV} \cr} \eqno(3.10)$$
$\rho_1$ as a function of $M_H$ is again plotted in Fig.3.
{}From Eq.(3.10) we see that the observable $O_{1,2}$ are much more
sensitive than $O_3$. The sensitivity for $a_{CP}$
is given simply by $y^{-1}$,
and it is about 30$\%$ larger than $\rho_1$.
\par\vskip 20pt
{\bf 4. Optimization} \par
In [9,10], it is discussed how to construct the most
sensitive observable
to the value of a physical parameter in a reaction, where
the distribution is linear in the physical parameter. We give here
briefly the result from [10]. Let us consider a scattering process
where the differential cross section has the form:
  $$ \Sigma (\phi) d\phi = (\Sigma_0(\phi) + g\cdot\Sigma_1(\phi))d\phi
      \eqno(4.1) $$
Here $\phi$ denotes the sample of the relevant phase-space variables,
with which we construct observables. The observable $T$
 which is most sensitive
to the parameter $g$ in Eq. (4.1) takes form:
  $$ T=T(\phi) = {\Sigma_1(\phi) \over \Sigma_0(\phi) } \eqno(4.2)$$
This result is generalized in [11] for more complicated cases. Such
types of the observables may be called optimal opservables,
following [11].
\par
In our case, with the distribution, i.e. the density matrix $R$ given
in Eq.(3.5) we construct according to Eq. (4.2)
the optimal observables for the form factors
in Eq.(3.2) which we want to measure. The observables are:
  $$ \eqalign { T_2&= { ({\bf\hat p_+}\cdot {\bf\hat q_+})
     ({\bf\hat q_+}\cdot {\bf\hat k}) M_Z E_Z
    +({\bf\hat q_+}\cdot {\bf\hat k})^2xM_Z(M_Z-E_Z) +x M_Z^2
    \over D_0 } \cr
      T_3&={-\beta_Z E_Z {\bf\hat q_+}\cdot
        ({\bf\hat p_+}\times{\bf\hat k}) (
      M_Z ({\bf\hat p_+}\cdot {\bf\hat q_+})
     +x(E_Z-M_Z)({\bf\hat q_+}\cdot {\bf\hat k})) \over D_0} \cr
    \ D_0&=E_Z^2(1-\beta_Z^2 x^2) +M_Z^2(
    {\bf\hat p_+}\cdot {\bf\hat q_+})^2 + 2M_Z(E_Z-M_Z)x
     ({\bf\hat p_+}\cdot {\bf\hat q_+})
     ({\bf\hat q_+}\cdot {\bf\hat k}) \cr
   & \ \ \ \ \  +({\bf\hat q_+}\cdot {\bf\hat k})^2
              (-\beta_Z^2E_Z^2 +x^2 (E_Z-M_Z)^2 \cr}
      \eqno(4.3) $$
The optimal observables $T_2$ and $T_3$ correspond to the simple observables
$O_2$ and $O_3$. To present the numerical results we define in anaolgy
to $w_i$ and $\rho_i$ the quantities:
  $$ <T_i>=z_i {B_t \over d}, \ \ \ \rho_{T_i}={\sqrt{ <T_i^2>}
    \over \vert z_i \vert } \eqno(4.4) $$
The numerical results are:
 $$ \eqalign{ z_2&=4.2\times 10^{-3},\ \ z_3=1.0\times 10^{-3}, \ \
    \rho_{T_2}=67,\ \ \rho_{T_3}=401 \ {\rm for\ } M_H=100{\rm GeV} \cr
    z_2&=4.2\times 10^{-3},\ \ z_3=9.3\times 10^{-4}, \ \
    \rho_{T_2}=60,\ \ \rho_{T_3}=414 \ {\rm for\ } M_H=200{\rm GeV} \cr
   } \eqno(4.5) $$
Comparing the sensitivity given in Eq.(3.10) for the simple observables
the sensitivity of the optimal observables
is from $20\%$ to 40$\%$ better.
However, the optimal observable $T_3$ is much less sensitive
than $T_2$, and even much
less sensitive than $O_{1,2}$.
\par
To construct the optimal observable
 $T_1$ corresponding to $O_1$, we integrate out the solid angle
of the letpton momentum, i.e. extract $A({\bf p_+},{\bf k})$ in
Eq. (2.3):
  $$ \eqalign { A({\bf p_+},{\bf k})& =({e^2 \over \sin^2\theta_W
    \cos^2\theta_W })^2 {d^2\over 32} {E_Z^2 s\over (s-M_Z^2)^2}
        \cdot \bigg\{ 2-\beta_Z^2 -\beta_Z^2 x^2
 \cr
 & + 8(1-{M_Z^2\over s})\sin\theta_W\cos\theta_W
      {M_Z\beta_Z \sqrt{s} \over E_Z}{\rm Im}F_\gamma(s)\cdot x \bigg\}
     \cr } \eqno(4.6) $$
In this simple case we find that the effect of the
improvement with the optimal observable is tiny, as also claimed
in [10]. Hence we do not present our numerical results for $T_1$.
\par\vskip 20pt
{\bf 5. Conclusion} \par
In this work we studied CP violating effects from the two Higgs doublet
model in $e^+e^-\rightarrow ZH$. The effects depend on a unknown
parameter ${B_t \over d}$ of the model, defined in Eq. (3.1).
To detect these effects we propose two sets of CP odd observables. One set
contains the simple CP odd observables, the other one contains
the so-called optimal observables, which are designed specially to maximize
the CP violating effects  from the two Higgs doublet model, or
more generally from the form factors in Eq. (3.2).  The
sensitivity of these observables to the parameter ${B_t\over d}$
is estimated. From our results, however, the optimal observables
do not improve the sensitivity to the CP violating effects
very much, the improvement is about $20\%$ to $40\%$, the price for this
improvement is that the
optimal observables
take a more complicated form  than the simple observables. \par
One
of the simple observables, i.e. the forward-backward asymmetry $a_{CP}$
of the $Z$ boson or the corresponding mean value  $<O_1>$ should
be readily accessible in in experiments. In principle, since the lepton
momentum is not involved here, more events should be
avaliable for measuring
$O_1$ and $a_{CP}$ than for the other observables
$O_{2,3}$ and $T_{2,3}$. Comparing the results obtained
in the previous sections, we conclude that the observable $O_1$ or
$a_{CP}$ are at most sensitive to the CP odd parameter ${B_t \over d}$.
However, the other observables should also be
studied in experiments, because our conclusion is specific to
two Higgs doublet model, and it is possible in other possible
extensions of the standard model that
$<O_1>$ and $a_{CP}$ both vanish,
but others  do not vanish.
\par
Finally, we note that one can determine how large $B_t$ should
at least be, independent of the parameter $d$, to have a possiblly
observable effect, if one knows the luminosity of the
$e^+e^-$ collider. Assuming the luminosity to be 10${\rm fb}^{-1}$
per year and requiring CP violation to be such that the expectation
value of $O_1$ is larger than twice of its statistical error, then
we require for $M_H=100$GeV:
   $$ B_t >7.2 \eqno (5.1) $$
for the observablity of the effect.
\par
\vskip 20pt
{\bf Acknowledgment:}\par
This work is supported in part by the Australian Research Council.
\vfil
\eject
\centerline{\bf Reference}\par\noindent
[1] J.H.Christensen, J. Cronin, V.F. Fitch and R. Turlay, Phys. Rev. Lett.
13 (1964) 138
\par\noindent
[2] C. Jarlskog, Phys. Rev. Lett. 55 (1985) 1039, Z. Phys. C29
(1985) 491
\par\noindent
[3] A. Brandenburg, J.P. Ma and O. Nachtmann, Z. Phys. C55 (1992) 115
\par\noindent
[4] G.C. Branco and M.N. Rebelo, Phys. Lett. B160 (1985) 117
\par\noindent
\ \ \ J. Liu and L. Wolfenstein, Nucl. Phys. B289 (1987) 1
\par\noindent
\ \ \ S. Weinberg, Phys. Rev. D42 (1990) 860
\par\noindent
[5] C.R. Schmidt and M. Peskin, Phys. Rev. Lett. 69 (1992) 410
\par\noindent
\ \ \ W. Bernreuther and A. Brandenburg, Phys. Lett. B314 (1993) 104,
   Phys. Rev.
\par\noindent
\ \ \ D49 (1994) 4481
\par\noindent
[6] J.P. Ma and B.H.J. McKellar, Phys. Lett. B319 (1993) 533
\par\noindent
[7] B. Grzadkowski and J.F. Gunion, Phys. Lett. b294 (1992) 361
\par\noindent
[8] W. Berneuther, O. Nachtmann, P. Overmann and T. Schr\"oder,
Nucl. Phys. B388 (1992) 53 B406 (1993) 516(E)
\par\noindent
[9] X.G. He, B.H.J. McKellar and S. Pakvasa, Int. J. Mod. Phys.
   A4 (1989) 5011
\par\noindent
[10] D. Atwood and A. Soni, Phys. Rev. D45 (1992) 2405
\par\noindent
[11] M. Diehl and O. Nachtmann, Heidelberg--Preprint, HD--THEP--93--37
\par\noindent
[12] D. Chang, W.Y. Keung and I. Phlips, Phys. Rev. D48 (1993) 3225
\par\noindent
[13] F.Abe et al (CDF Collarboration) Preprint, FERMILAB-PUB-94/116-E
\par\noindent
\vskip 15pt
\centerline{Figure Caption}\par\vskip 15pt
Fig.1: One of the Feyman diagrams contributing to the form factor
 in Eq. (3.2) \par
Fig.2: The parameter $y$ for the CP asymmetry $a_{CP}$ as a function
of the Higgs mass. \par
Fig.3: The sensitivity $\rho_1$ for the observable $O_1$ as a function
of the Higgs mass.

\par\vfil
\end
\eject
\centerline{\bf Fig.1}
\par\vfil\eject
\centerline{\bf Fig.2}
\par\vfil\eject
\centerline{\bf Fig.3}
\par\vfil\eject
\end